\documentclass[twocolumn, pra, showpacs,superscriptaddress,floatfix]{revtex4}
\usepackage{graphicx}
\usepackage{dcolumn}
\usepackage{bm}
\usepackage{amsmath}

\begin{document}

\title{Transition from Tonks-Girardeau gas to super-Tonks-Girardeau gas as
an exact many-body dynamics problem}

\author{Shu Chen}
\affiliation {Institute of Physics,
Chinese Academy of Sciences, Beijing 100190, China}
\author{Liming Guan} \affiliation {Institute of Physics,
Chinese Academy of Sciences, Beijing 100190, China}
\author{Xiangguo
Yin} \affiliation {Institute of Physics, Chinese Academy of
Sciences, Beijing 100190, China}
\author{Yajiang Hao}
\affiliation{Department of Physics, University of Science and
Technology Beijing, Beijing 100083, China}
\author{Xi-Wen Guan}
\affiliation{Department of Theoretical Physics, Research School of
Physics and Engineering, Australian National University, Canberra
ACT 0200, Australia}

\begin{abstract}

We investigate transition of a one-dimensional interacting Bose gas
from a strongly repulsive regime to a strongly attractive regime,
where a stable highly excited state known as the super
Tonks-Girardeau gas was experimentally realized very recently. By
solving exact dynamics of the integrable Lieb-Liniger Bose gas, we
demonstrate that such an excited gas state can be a very stable
dynamic state. Furthermore we calculate the breathing mode of the
super Tonks-Girardeau gas which is found to be in good agreement
with experimental observation. Our results show that the highly
excited super Tonks-Girardeau gas phase can be well understood from
the fundamental theory of the solvable Bose gas.
\end{abstract}

\pacs{03.75.Kk, 05.30.Jp, 05.30.-d}
\date{\today}
\maketitle


{\it Introduction.---} Recent experimental advances with ultracold
atoms have opened up a new avenue for the study of one-dimensional
(1D) strongly correlated many-body systems which continue to inspire
significant developments in physics
\cite{gorlitz,esslinger,Paredes,Toshiya}. Exquisite tunability of 1D
effective interacting strength via Feshbach resonance or the
confinement-induced resonance \cite{Olshanii,Olshanii2,Sinha} allows
the simulation of low-dimensional quantum gases with arbitrary
interaction strength. The experimental realization of
Tonks-Girardeau (TG) gases \cite{Paredes,Toshiya} has provided a
direct test on the fermionization of Bose gas in the strongly
interacting limit \cite{Girardeau,Lieb}.
Very recently, Haller {\it et al} \cite{Haller} have made a new
experimental breakthrough on the realization of the stable highly
excited gas-like phase called the super-Tonks-Girardeau (STG) gas,
which was predicted theoretically \cite{Astrakharchik1} and verified
from the integrable interacting Bose gas with attractive
interactions \cite{Batchelor}. Their experimental results have
stimulated further theoretical study of such novel states with
strong correlations in 1D many-body systems \cite{Girardeau-new},
which was also found theoretical correspondence in a strongly
attractive Fermi gas without polarization \cite{Chen}.

In general, experimental realization of a stable excited state is
difficult and a little counter-intuition since it has no analogue in
 traditional solid state systems, where a pure excited state is not
stable due to the energy dissipation between the system and the
environment. 
The experimental progress \cite{Haller} opens the door for searching
novel quantum states in quantum gases and studying physical
properties of stable excited phases, which provides a promising new
area of activity in cold atoms. However, the metastability, dynamics
and statistical signature of the excited states are still far from
understood. Current theoretical understanding of the STG gas is
based on intuitive explanation that the STG gas inherits  hard core
behavior from the repulsive TG gas which prevents the collapse to
its true cluster ground state (GS)
\cite{Batchelor,Astrakharchik1,Chen,McGuire,Astrakharchik2,Tempfli,Girardeau-new}.
As a matter of fact, the STG gas is obtained from the TG gas by a
sudden switch of interactions, this process is in principle a
dynamics problem for the interacting boson model.
Therefore a comprehensive understanding of the STG gas is highly
desirable and it is also important to show that such a stable state
really follows from the fundamental quantum dynamics of an exactly
solvable model.

In this work, we demonstrate how the GS of the repulsive TG gas
translates to the highly excited state of the strongly attractive TG
gas through a sudden switch of interactions. We also show that the
highly excited STG phase is stable and the transition rate to the GS
is almost completely suppressed in the strongly interacting regime.
Furthermore, we calculate the energy of the STG gas by solving the
exact Bethe ansatz equations (BAEs) and determine the collective
excitation modes of the STG gas with the help of the local density
approximation (LDA), which agrees quantitatively  with the
experimental data.

{\it Interacting boson model.---} We consider a quasi-1D system
with $N$ bosons tightly confined in an elongated trap which is
described by an effective 1D Hamiltonian
\begin{equation}
H=\sum_{i=1}^N -\frac{\hbar ^2}{2m}\frac{\partial ^2}{\partial x_i^2}%
 + g\sum_{i<j}\delta (x_i-x_j), \label{H1}
\end{equation}
with $g =- 2\hbar ^2/(m a_{1d})$ being the effective 1D interaction
strength and $a_{1d}$ the effective 1D scattering length \cite
{Olshanii,Olshanii2}.
As a simple but fundamental integrable interacting boson model, the
exact results for the Lieb-Liniger (LL) model (\ref{H1}) play an
important role in understanding quantum statistical and many-body
correlation effects of a 1D bosonic quantum gas
\cite{Lieb,Girardeau03,Buljan,Gangardt,Dunjko}.
The model (\ref{H1}) is exactly solved by the Bethe ansatz method
with eigenfunctions taken as superpositions of plane waves over all
permutations of the momenta
\begin{equation}
\Psi(x_1,\cdots,x_N) = \sum_P A_P e^{\mathrm{i}\sum_j k_{P_j} x_j},
\label{wave}
\end{equation}
where the wave functions are defined in the domain $x_1
<x_2<\cdots<x_N$ and the coefficients $ A_P=(-1)^P \prod_{j<l}^N
\frac{i k_{p_l}-i k_{p_j} + c} {\sqrt{(k_{p_l}-k_{p_j})^2+c^2}} $
are functions of the two-particle scattering phase shifts with
$(-1)^P=\pm 1$ for odd or even $P$. The quasi-momenta $k_j$ are
determined by the BAEs \cite{Lieb}
\begin{equation}
\exp\left( \mathrm{i}k_{j}L\right)  = - \prod_{l=1}^{N}\left( \frac{
k_{j}-k_{l}+\mathrm{i}c}{k_{j}-k_{l}-\mathrm{i}c} \right). \label{BAEbose}
\end{equation}
Here the coupling constant $g ={\hbar ^2 c}/{m}$ with interaction
strength $c=-{2}/{a_{\rm 1d}}$ determined by the effective 1D
scattering length $a_{\rm 1d}$.  The total momenta of the system is
given by $K=\sum_{j=1}^{N} k_j$ and the eigenenergies are given by $
E=\frac{\hbar^2}{2m} \sum_{j=1}^{N} k_j^2 $.

\begin{figure}[tbp]
\includegraphics[width=8.2cm]{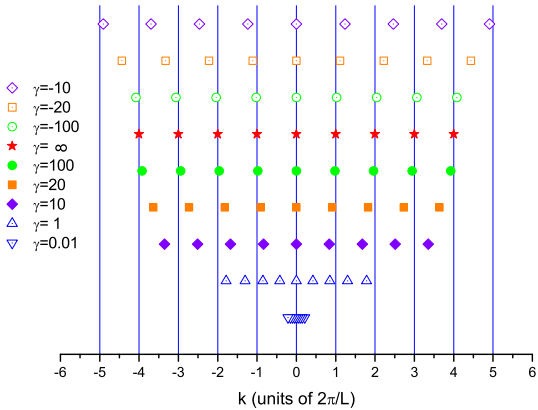}
\caption{ (color online) Quasi-momentum distributions for the ground
state of the repulsive and the STG gas phase of the attractive Bose
gas with different values of $\gamma$.} \label{momentum}
\end{figure}

For repulsive interaction $c>0$, all solutions $k_j$ of
(\ref{BAEbose}) with $j=1,\cdots,N$ are real numbers. Alternatively
the solutions follow by solving the log form
\begin{equation}
k_j L= {2\pi} I_{j} - \sum_{i =1}^{N} 2 \arctan \left[
(k_{j}-k_{i})/{\left\vert c \right\vert } \right] \label{BAETG}
\end{equation}
of the BAEs. The GS solution corresponds to $I_j=(N+1)/2-j$. On the
other hand, for attractive interaction $c<0$, the GS solution to the
BAEs follows from the $N$-string solution $k_{j}=\lambda
_{j}+\mathrm{i}(N+1-2j) (c/2 + \delta_j)$, where $j=1, \cdots ,N$
and $\delta_j$ are small deviations which fall off exponentially to
zero as $c \rightarrow -\infty$. Correspondingly, the GS is
described by a cluster state \cite{McGuire} with the energy
$E_0=-\frac{1}{12} c^2 N(N^2-1)$.

{\it STG phase---} Although the GS solution corresponds to the
$N$-string solution,  the BAEs (\ref{BAEbose}) still have real
solutions even for $c<0$, which however correspond to some highly
excited state of the attractive Bose gas and can be determined by
solving the BAEs
\begin{equation}
k_j L= {2\pi} I_{j} + \sum_{i =1}^{N} 2 \arctan \left[
(k_{j}-k_{i})/{\left\vert c \right\vert } \right] . \label{BAEsTG}
\end{equation}
For the solutions with $I_j=(N+1)/2-j$, the corresponding state is
just the STG gas state \cite{Batchelor}. In the strongly attractive
interaction limit ($c \rightarrow -\infty$), the quasi-momentum
distribution of real roots determined from (\ref{BAEsTG}) reduces to
the free Fermi distribution with a minimum separation between two
quasimomenta $2\pi/L$. As the attraction strength decreases, the
gas-like highly excited state has a stronger anyonic-like pressure
than the Fermi pressure. For strongly attractive interaction ($c
\rightarrow -\infty$), the  energy of the STG gas state
\cite{Batchelor} is given by
$ E_{STG} = \frac{\hbar^2}{2m} \frac{1}{3} N \left(
N^{2}-1\right)\frac{\pi ^{2}}{L^{2}}\left[ 1 + \frac{4
N}{L\left\vert c \right\vert} + \frac{12 N^2}{L^2\left\vert c
\right\vert^2} + \cdots \right] $ , which is continuously connected
to the GS energy of the TG gas
$ E_{TG} = \frac{\hbar^2}{2m} \frac{1}{3} N \left(
N^{2}-1\right)\frac{\pi ^{2}}{L^{2}}\left[ 1 - \frac{4
N}{L\left\vert c \right\vert} + \frac{12 N^2}{L^2\left\vert c
\right\vert ^2} - \cdots \right] $ as $|c|\rightarrow \infty$
\cite{Girardeau}.

{\it Quench dynamics---} Suppose that the initial state $|
\Psi_{in}(X,t=0) \rangle = | \psi_0(X,c) \rangle$ is prepared in the
strongly repulsive limit $c \gg 1$, after a quick switch from the
repulsive regime to the attractive regime with $c'<0$, the
wavefunction $| \Psi(X,t)\rangle = e^{\mathrm{i}Ht} | \Psi_i(X,c)
\rangle$ can be calculated via
\begin{equation}
| \Psi(X,t)\rangle = \sum_{n=0}^{\infty} e^{i E_n t} c_n | \psi_n
(X,c')\rangle , \label{Psit}
\end{equation}
where $c_n = \langle \psi_n(X,c')| \psi_0(X,c)\rangle$ with
$\psi_n(X,c')$ representing the $n$-th eigenstate of the LL model
with attractive interaction strength $c'$ and $X$ the abbreviation
of $x_1,\cdots,x_N$. From (\ref{Psit}), it is straightforward that
the probability for the  state after quench staying in a STG phase
is given by $|\langle \psi_{STG}(X,c')| \psi_0(X,c)\rangle|^2$,
which also represents the transition probability from an initial TG
phase to the final STG phase. The wavefunction $\psi_{STG}$ is
determined by the Bethe ansatz wave function (\ref{wave}) with $k_j$
determined by the solutions of (\ref{BAEsTG}) with $I_j=(N+1)/2-j$.
Similarly, the transition probability from the initial TG state to
the cluster state is given by $|\langle \psi_{cluster}(X,c')|
\psi_0(X,c)\rangle|^2$. Since the Bethe ansatz wave functions
$\psi_n(X,c)$ can be exactly determined by solving BAEs, in
principle we can calculate the transition probabilities exactly.

For simplicity, we first consider the case with $c'=-c$, i.e., the
system is initially in the GS of the LL gas with $c>0$, and then
suddenly switch to the attractive regime with strength $-c$. We note
that the BAE solutions for the repulsive LL gas and that for the STG
gas correspond to the same set of ${I_j}$ according to Eqs.
(\ref{BAETG}) and (\ref{BAEsTG}). In the limit $c=-c' \rightarrow
\infty$, the solutions are given by $k_j = I_j 2 \pi/L$, {\it i.e.}
they are exactly the same. In Fig. 1, for particle number $N=9$, we
show the BAE solutions for the repulsive LL gas and the STG gas with
different values of $\gamma=c/\rho$. In the strongly interacting
regime, i.e. $|\gamma| \gg 1$, the quasi-momentum distributions for
the repulsive LL gas and the STG gas approach the free fermion
``orbitals'' from different sides. As $|\gamma|$ decreases, the
deviation from the free-fermion distribution become more obvious.
From the change of momentum distribution, we can expect that the
overlap between the GS wavefunction of the strongly repulsive LL gas
and the eigen-wavefunction for a STG gas, i.e., $|\langle
\psi_{STG}(X,-c)| \psi_0(X,c)\rangle|$, is close to $1$ for very
large $|\gamma|$ and exactly  $1$ when $|\gamma| \rightarrow
\infty$.

To see how the transition rate changes with respect to $\gamma$, we
calculate the overlap of the wavefunctions between the repulsive and
attractive regimes. In order to conceive the signature of the
overlap between the two regimes, we consider a system with $N=4$ and
calculate the transition rate from the initial repulsive LL gas to
the STG gas with different $|\gamma|$. In Fig. 2, we show the
transition rate from an initial repulsive gas with $\gamma=200$, to
a final STG phase with different values $\gamma<0$. As expected, the
transition probability to the STG phase is close to $1$ after
switching to the strongly attractive regime whereas the probability
for dynamically falling into the cluster state is almost completely
suppressed. However, when the system is switched into the weakly
attractive regime, the STG phase is no longer stable and the
probability for falling into the cluster state increases quickly.
The transition rate for the larger system does not change
qualitatively, however the calculation for a large system is a very
time-consuming task due to the calculation of multidimensional
integrals.
Our calculations give a clear signature of the metastable STG gas
against collapsing into the cluster state.
\begin{figure}[tbp]
\includegraphics[width=8.2cm,angle=-0]{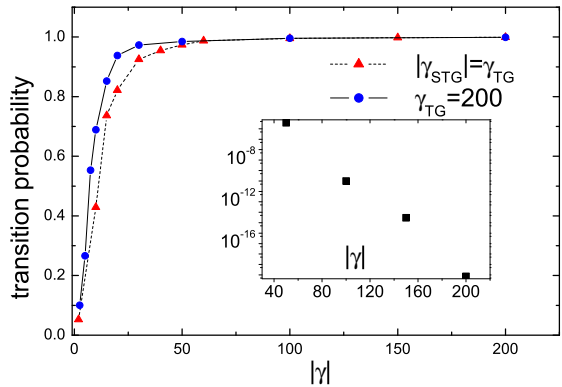}
\caption{ (color online) Transition probabilities from the TG gas to
STG phase. Inset: the transition probabilities from TG gas to the
cluster state.} \label{transition}
\end{figure}
\begin{figure}
\includegraphics[width=8.20cm]{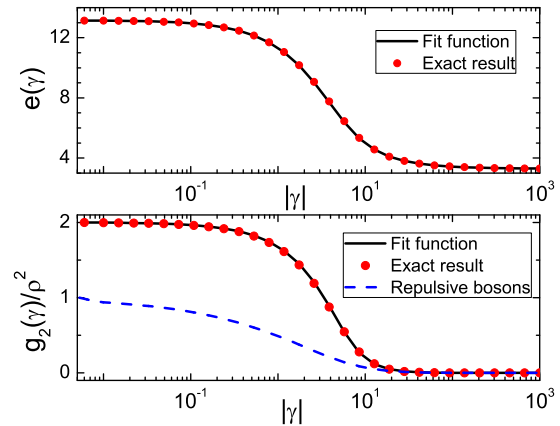}
\caption{$e(\gamma)$ and $g_2(\gamma)$ versus $|\gamma|$.}
\end{figure}

{\it Properties for the STG gas---} As shown in the above
calculation, the transition rate to the STG phase is very high in
the strongly interacting regime. Therefore such a highly excited
state with a stronger pressure than the Fermi pressure could be
reached through switching the interaction from strong repulsion into
strong attraction. We note that such pressure is inherited from the
repulsive gas, i.e. transmutes the statistical kinetic energy into
the attractive STG phase. In the thermodynamic limit with $N,L
\rightarrow \infty$, the eigen energy of the STG phase can be
represented in  the integral form: $E_{STG}/L = \frac{\hbar^2}{2m}
\int k^2 \rho(k) \frac{d k}{2 \pi}$ and $ \rho(k) = \frac{1}{2\pi}-
\frac{1}{2\pi}\int_{-Q}^{Q}\frac{2 \left \vert c
\right\vert\rho(k^{\prime}) }{c^2+(k-k^{\prime})^2}dk^{\prime}$,
where the integration limit $Q$ is determined by $\rho=\int_{-Q}^Q
\rho(k)dk$. By making replacement of variables $k=Qx$, $c=Q\lambda$
and $ g(x)=\rho(Q x)$ according to \cite{Lieb}, the energy per atom
can be expressed as $ \epsilon \left( \rho \right) =\frac{\hbar
^2}{2m} \rho ^2 e\left( \gamma \right)$, which can be obtained by
solving the system of equations
\begin{eqnarray}
e\left( \gamma \right) &=& \left| \frac{\gamma}{\lambda} \right|^3
\int_{-1}^{1}
g(x) x^2 d x , \label{LL1} \\
g(x) &=& \frac{1}{2\pi}-\frac{1}{2\pi}\int_{-1}^{1}\frac{2 \lambda
g(x^{\prime})}{\lambda ^2+(x-x^{\prime})^2}d x^{\prime}, \label{LL2}
\end{eqnarray}
and $\lambda = |\gamma| \int_{-1}^1 g(x) dx $. The above equations
are very similar to Lieb-Liniger's solutions for the repulsive boson
gas except the minus sign in Eq. (\ref{LL2}).
\begin{figure}[tbp]
\includegraphics[width=8.20cm]{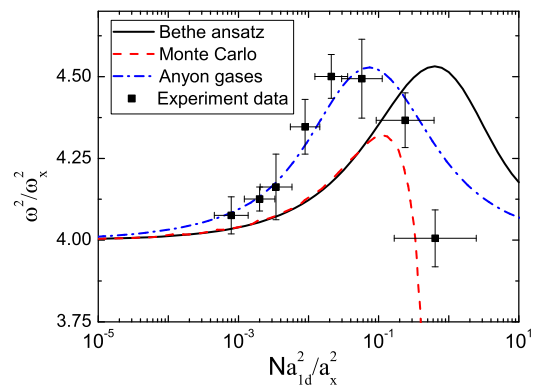}
\caption{ (color online) Breathing mode of the STG gas. The
experiment data with error bars and Monte Carlo result are
reproduced from Fig. 3a of Ref. \cite{Haller}. \label{breathing
mode}}
\end{figure}
By numerically solving the integral equation, we can obtain
$e(\gamma)$ and thus the energy density $\epsilon \left( \rho
\right) $ and the chemical potential $\mu(\rho)=
\partial_{\rho}[\rho \epsilon(\rho) ]$. For the STG phase in the
strongly attractive regime, we find $ e( \gamma )\approx
\frac{\pi^2}{3}\left( 1+\frac{4}{|\gamma|}+\frac{12}{|\gamma|^2}+
\frac{32}{|\gamma|^3}\left(1-\frac{\pi^2}{15}\right)\right)$.

In Fig.3, we show $e(\gamma)$ for different values of $\gamma$,
which is found to be well fitted in the whole attractive regime by a
rational function $e\left(  \gamma\right)
=\frac{4\pi^{2}}{3}(1+p_{1}\left\vert
\gamma\right\vert +p_{2}\gamma^{2}+p\left\vert \gamma\right\vert ^{3}%
/4)/(1+q_{1}\left\vert \gamma\right\vert
+q_{2}\gamma^{2}+p\left\vert \gamma\right\vert ^{3})$ with
$p_1=0.075$, $p_2=0.013$, $q_1=0.227$, $q_2=0.034$, $p=0.004$.
We also calculate the local two-particle correlation function
$g_2=\langle \Psi^{\dagger}\Psi^{\dagger}\Psi\Psi\rangle$ which can
be obtained by $g_2(\gamma)=\rho^2 d e(\gamma)/d \gamma$
\cite{Gangardt}. The STG phase exhibits stronger local correlation
than the repulsive Bose gas. For the STG gas in a harmonic trap with
$V_{ext}=m\omega_x^2x^2/2$, we can determine the density
distribution of the STG gas using LDA \cite {Dunjko,Menotti}.

According to the LDA, one can assume that the system is in local
equilibrium at each point $x$ in the external trap.
The density distribution of the STG gas is
then obtained from the local equation of state
$\mu[\rho(x)]=\mu_0-V_{ext}(x)$ under Thomas-Fermi approximation
with the normalization condition $\int dx \rho(x)=N$. Using the
density distribution, one can then calculate the mean square radius
of the trapped STG gas via $\langle x^2 \rangle = \int \rho(x) x^2
dx/N$. Following Ref. \cite{Menotti,Astrakharchik1}, we can
calculate the frequency of the lowest breathing mode given by $
\omega^2=-2 \langle x^2 \rangle/(d\langle x^2 \rangle/d\omega_x^2)$.
Comparing with the Monte Carlo simulations \cite{Astrakharchik1},
our result (solid line) based on the exact solutions of BAEs has a
pronounced peak with a maximum of breathing mode frequency
$\omega^2/\omega_x^2$ reaching the value about $4.5$ which coincides
with the experimental data, see Fig~ \ref{breathing mode}.  Here the
discrepancy between the theoretical prediction from the LL model and
the experimental data
inspires us to search for a new theoretical framework capable of
understanding the quantum dynamics of the excited quantum gas phase.

To this goal, we notice that the resonance regime of switching the
interaction from strong repulsion to strong attraction can be
identified as anyonic type of interaction $c/\cos(\kappa/2)$
\cite{Kundu,BGO}. We also understand that strongly interacting
bosons have lost their bosonic signature due to the transmutation
between statistical and dynamical interactions \cite{Chen}.  This
naturally suggests us to consider the integrable model of anyons as
a fitting model to the experiment on the STG gas. The dynamic
process of switching the interaction from strong repulsion to strong
attraction may be viewed as the change of quantum statistics from
less to more exclusive than Fermi statistics. In order to conceive
this nature of the STG gas, we also calculate the breathing mode of
the integrable model of anyons with $\kappa=3.93$ greater than the
free Fermi case $\kappa=\pi$, see dash-dotted line. The good
agreement with experiment data implies that the anyonic signature
may relate to the statistical signature of the highly excited state
of interacting bosons. However, we notice that in the weak coupling
regime the involvement of cluster states of few particles
\cite{Girardeau-new} is the main reason for the discrepancy between
theoretical prediction and experimental data. When the system is not
in the strongly attractive regime, the STG phase is no longer
stable, see Fig~\ref{breathing mode}.

{\it Summary---} In summary, we have investigated the transition
from the GS of strongly repulsive bosons to the gas-like highly
excited state of attractive bosons through a switch of interaction.
By solving the quench dynamics problem of the interacting bosons, we
have shown that the gas-like excited state is stable in the strongly
attractive regime. Our exact result for the frequency of the lowest
breathing mode obtained from the BAEs for interacting bosons and
anyons are in reasonably good agreement with experimental
observation. This suggests a statistical signature of the STG gas
which may be viewed  as an anyon-like gas.

This work is supported by NSFC under Grants Nos. 10821403 and
10974234, 973 project (China). XWG has been supported by the
Australian Research Council. We thank M T Batchelor for helpful
discussions and critical reading of the paper.


\begin{thebibliography}{99}
\bibitem{gorlitz}  A. G\"{o}rlitz, et.al., Phys. Rev. Lett. \textbf{87},
130402 (2001).

\bibitem{esslinger}  H. Moritz, et.al., Phys. Rev. Lett. \textbf{91}, 250402 (2003);
T. St\"{o}ferle, et.al., \textsl{ibid.} \textbf{92}, 130403 (2004).




\bibitem{Paredes}  B. Paredes, {\it et. al.}, Nature \textbf{429}, 277 (2004).


\bibitem{Toshiya}  T. Kinoshita, {\it et. al.}, Science \textbf{
\ 305}, 1125 (2004).



\bibitem{Olshanii}  M. Olshanii, Phys. Rev. Lett. \textbf{81}, 938
  (1998).

\bibitem{Olshanii2} T. Bergeman, {\it et. al.}, Phys. Rev.
Lett. \textbf{91}, 163201 (2003).

\bibitem{Sinha}S. Sinha {\it et. al.}, Phys. Rev. Lett. \textbf{99}, 140406 (2007).

\bibitem{Girardeau} M. D. Girardeau, J. Math. Phys. (N.Y.) \textbf{1}, 516
(1960).

\bibitem{Lieb}  E. H. Lieb and W. Liniger, Phys. Rev. \textbf{130}, 1605
(1963).

\bibitem{Haller} E. Haller, {\it et. al.}, Science \textbf{325}, 1224 (2009).

\bibitem{Astrakharchik1} G. E. Astrakharchik, J. Boronat, J. Casulleras, and S.
Giorgini, Phys. Rev. Lett. \textbf{95}, 190407 (2005).

\bibitem{Batchelor} M. T. Batchelor, M. Bortz, X. W. Guan, N.
Oelkers, J. Stat. Mech. (2005) L10001.

\bibitem{Girardeau-new} M. D. Girardeau {\it et. al.}, arXiv: 0912.1633.

\bibitem{Chen} S. Chen, {\it et. al.}, arXiv: 0910.1652.

\bibitem{McGuire} J. B. McGuire, J. Math. Phys. (N.Y.) \textbf{5}, 622
(1964).

\bibitem{Astrakharchik2} G. E. Astrakharchik, D. Blume , S. Giorgini and  B. E. Granger, Phys.
Rev. Lett. \textbf{92}, 030402 (2004).

\bibitem{Tempfli} E. Tempfli, {\it et. al.}, New J. Phys. 10, 103021 (2008).





\bibitem{Girardeau03} M. D. Girardeau, Phys. Rev. Lett. \textbf{91}, 040401 (2003).


\bibitem{Buljan} H. Buljan, {\it et. al.}, Phys. Rev. Lett. \textbf{100}, 080406 (2008).







\bibitem{Gangardt} D. M. Gangardt and G. V. Shlyapnikov, Phys. Rev.
Lett. {\bf 90}, 010401 (2003).


\bibitem{Dunjko} V. Dunjko, {\it et. al.}, Phys. Rev. Lett. {\bf 86}, 5413 (2001).

\bibitem{Menotti} C. Menotti, {\it et. al.}, Phys. Rev. A \textbf{66},
043610 (2002).

\bibitem{Kundu} A. Kundu, {Phys. Rev. Lett.} {\bf 83}, 1275 (1999).
\bibitem{BGO}M. T. Batchelor, {\it et. al.}, Phys. Rev. Lett. {\bf
96}, 210402 (2006).

\end{thebibliography}
\end{document}